\documentclass[conference]{IEEEtran}

\usepackage{cite}
\usepackage{url, color, soul}      	
\usepackage{balance}  				
\usepackage{graphicx} 				
\usepackage[]{algorithm2e}
\usepackage{booktabs}
\usepackage{subfig}
\usepackage{xspace}
\usepackage{multirow}
\usepackage{tabularx}
\usepackage{pbox}
\usepackage[cmex10]{amsmath}
\usepackage{bm}

\ifCLASSINFOpdf
  \else
  
\fi
\newcommand{\todo}[1]{}

\begin{document}

\title{Spotting Suspicious Reviews via \\(Quasi-)clique Extraction}

\author{
    \IEEEauthorblockN{Paras Jain, Shang-Tse Chen, Mozhgan Azimpourkivi$^\dagger$,
    Duen Horng Chau, Bogdan Carbunar$^\dagger$}
    \IEEEauthorblockA{Georgia Tech, $^\dagger$Florida International University\\
    \{paras, schen351, polo\}@gatech.edu, $^\dagger$\{mazim003, carbunar\}@cs.fiu.edu
	}
}

\maketitle

\begin{abstract}
How to tell if a review is real or fake?
What does the underworld of fraudulent reviewing look like?
Detecting suspicious reviews has become a major issue for many online services. 
We propose the use of a clique-finding approach to discover well-organized suspicious reviewers. 
From a Yelp dataset with over one million reviews,
we construct multiple \textit{Reviewer Similarity} graphs to link users that have unusually similar behavior:
two reviewers are connected in the graph if they have reviewed the same set of venues within a few days.
From these graphs, our algorithms extracted many large cliques and quasi-cliques, the largest one containing a striking 11 users who coordinated their review activities in identical ways.
Among the detected cliques, a large portion contain \textit{Yelp Scouts} who are paid by Yelp to review venues in new areas.
Our work sheds light on their little-known operation.
\end{abstract}

\section{Introduction}

Review-centric online services like Yelp\footnote{\url{http://www.yelp.com}} and TripAdvisor crowdsource the job of reviewing businesses. 
The popularity and influence of reviews make such sites ideal targets for malicious behaviors: businesses commission fraudulent reviews to artificially boost their ratings. 
An estimated 16\% of Yelp restaurant reviews are fraudulent \cite{Luca2014}.

Identifying suspicious review behaviors is critical to maintaining the integrity of online services and protecting their users. 
However, this task is challenging, as fraudsters' strategies can change rapidly. Crowdsourcing services such as Freelancer, Fiverr and Amazon Mechanical Turk are exploited for recruiting experienced review writers at a massive scale for nefarious purposes \cite{fiverrmturk}.

Recent research started to investigate network-based techniques for uncovering organized fraud by analyzing the link structures among potential fraudsters. 
For example, NetProbe uses an inference algorithm to find ``near bipartite cores'' formed among fraudsters and their accomplices on eBay \cite{chau2007netprobe}. 

More recently, Vlasselaer et al. find rectangles in bipartite graphs to detect social security fraud \cite{van2015guilt}.

Interestingly, even though cliques\footnote{A complete sub-graph} and quasi-cliques\footnote{Synonymous with \textit{pseudo cliques}} have long been hinted as one of the strongest tell-tale signs of fraud, no prior work has studied if they indeed exist in online review websites like Yelp,
where we can create a graph where each node represents a user, and an edge connects two users if they have reviewed common venues. 

Our first research goal is to mine such graphs for cliques and quasi-cliques to verify the hypotheses from literature.
Our secondary goal is to study if such cliques are indeed suspicious. 
Or relatedly, whether the (quasi-)clique structure is a strong indicator for fraud, whether there may be false convictions (e.g., any ``good'' cliques?), and if so, whether they are common and what approaches can reduce them.

\begin{figure}[t]
\centering
\includegraphics[width=0.9\linewidth]{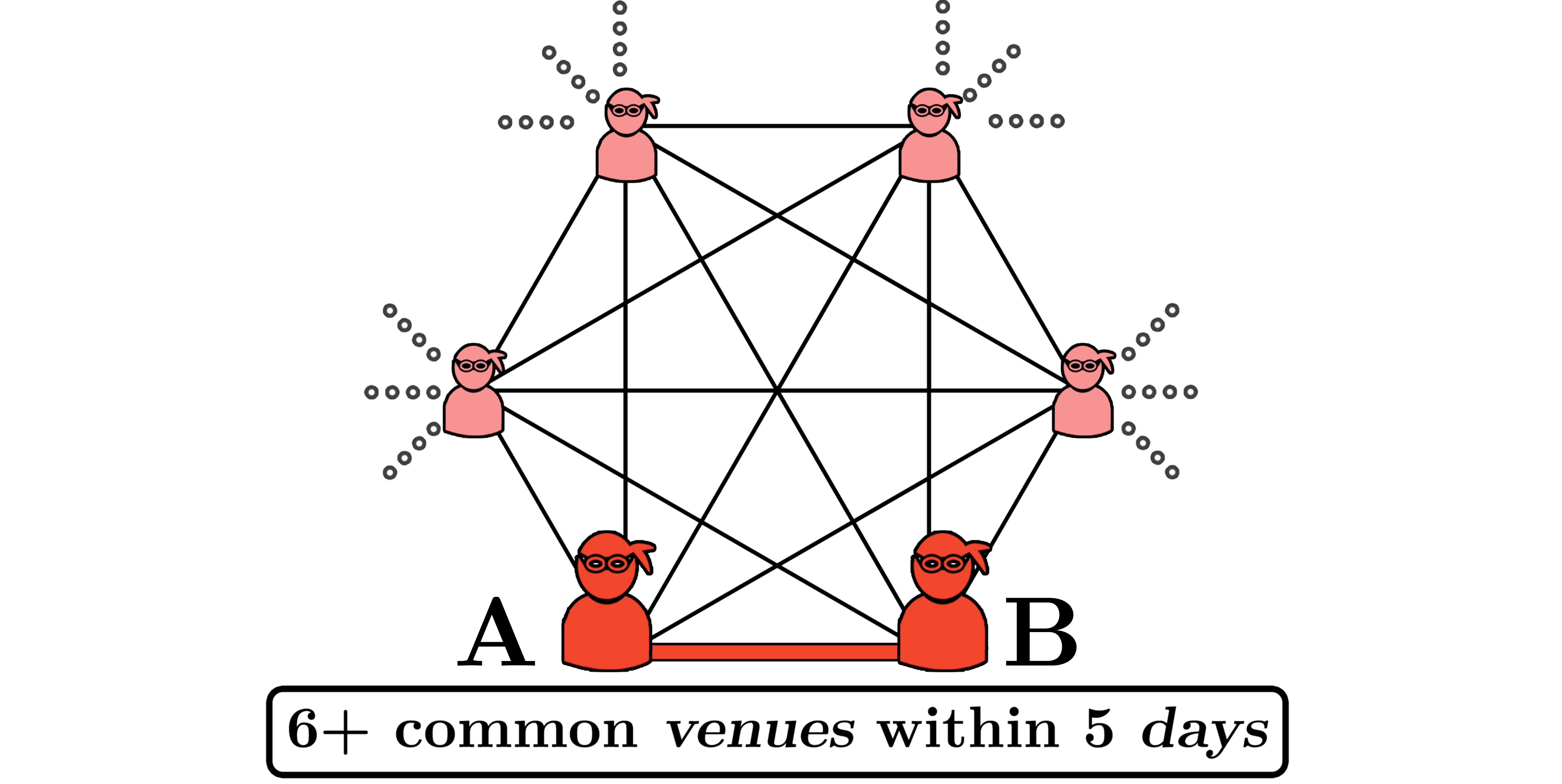}
\caption{Example suspicious clique formed among reviewers, found in a $(6, 5)$-graph extracted from the Yelp data.
Two reviewers are linked when they have reviewed the same 6 venues (or more) within 5 days.}
\label{fig:cliquechart}
\end{figure}

We describe our preliminary results, which show that even from public data provided by Yelp, we can find large cliques that involve as many as 11 users---intuitively, this means that every possible pair of users (among these 11) reviewed multiple common venues within only a few days apart (see Figure \ref{fig:cliquechart}).
In practice, we should rarely see cliques of such large sizes. 
Possibly, the only legitimate setting for that to happen is that those 11 people are close friends or family members who always go to the same places together and also write reviews together!

\section{Extracting (Quasi-)Cliques}

\noindent\textbf{The Yelp Review Dataset.}
We use the dataset from the \textit{Yelp Dataset Challenge} \footnote{\url{http://www.yelp.com/dataset_challenge}}, which contains 42,153 venues and 1,125,458 reviews. Yelp did not specify whether they filtered the dataset but it would be reasonable to assume that they have only included publicly listed reviews in the dataset.

\noindent\textbf{Building K-D Graphs to Uncover Suspicious Links.} 
We extract a set of $(k, d)$-graphs from the raw Yelp data, originally formatted as a list of JSON objects, varying the $k$ and $d$ parameters. We define a $(k, d)$-graph to be an undirected graph, where vertices represent users, and an edge\footnote{\label{weightededge}Edges can be weighted with a calculated similarity score between users} exists between two vertices if the corresponding users reviewed at least $k$ venues in common, and the reviews for each venue were posted at most $d$ days apart (see Table \ref{table:graphfeatures}).
\renewcommand\arraystretch{1.8}
\begin{table}[t]
    \begin{tabularx}{\linewidth}{l X l}
    
    \toprule
   
    \textbf{\textit{k}} & Minimum number of commonly venues reviewed (by two users) \\[-0.6em]
    \textbf{\textit{d}} & Maximum number of days between two reviews (written for the same venue) \\[-0.6em]
	\textbf{Node}      & A Yelp Reviewer\\[-0.6em]
    \textbf{Edge}      & Connects two users who reviewed $\geq k$ same venues within $d$ days \\[-0.1em]
    \bottomrule
	\end{tabularx}
    \caption{$(k,d)$-graph definition (Reviewer Similarity Graph).}
    \label{table:graphfeatures}
\end{table}
\renewcommand\arraystretch{1}

\begin{table}[b]
\small
\centering
\begin{tabularx}{\columnwidth}{l c c c c c c c c c}
\toprule
\textbf{\textit{(k, d)}-graph} & 3,5 & 3,6 & 3,8 & 4,5 & 4,6 & 5,5 & 5,6 & 6,5 \\ \midrule
\textbf{9--clique}  & 112 & 152 & 1040				 		   & 29  & 73 						   & 13 & 28 & 10 \\
\textbf{10--clique} & 22  & 25  & 290 				 		   & 3   & 13 						   & 1  & 3  & \textcolor{red}{\textbf{1}}  \\
\textbf{11--clique} & \textcolor{red}{\textbf{2}}   & \textcolor{red}{\textbf{2}}   & \textcolor{red}{\textbf{50}} & ---   & \textcolor{red}{\textbf{1}} & ---  & ---  & ---  \\

\bottomrule
\end{tabularx}
\caption{Counts of large suspicious cliques, of sizes 9, 10, and 11, found in select $(k, d)$-graphs\protect\footnotemark. The most suspicious cliques are highlighted in red, due to large sizes, higher $k$ (more venues) and lower $d$ values (tighter time bound).}
\label{table:cliqueresults}
\end{table}

\begin{table}[b]
\small
\centering
\begin{tabularx}{\columnwidth}{l c c c c c c c}
\toprule
\textbf{\textit{(k, d)}-graph} 	   & 6,5 & 6,8  & 7,5 & 7,8  & 8,5 & 8,8 & 9,5 \\ \midrule

\textbf{9--quasiclique}    & 144 & 649  & 94  & 351  & 42  & 227 & 8   \\
\textbf{10--quasiclique}   & 44  & 315	& 33  & 134  & 12  & 84  & --- \\
\textbf{11--quasiclique}   & \textcolor{red}{\textbf{7}} & \textcolor{red}{\textbf{100}} & \textcolor{red}{\textbf{4}} & \textcolor{red}{\textbf{33}}	& --- & \textcolor{red}{\textbf{15}} & --- \\
\textbf{12--quasiclique}   & \textcolor{red}{\textbf{1}} & \textcolor{red}{\textbf{20}} & --- & \textcolor{red}{\textbf{4}} & --- & \textcolor{red}{\textbf{1}} & --- \\
\bottomrule
\end{tabularx}
\caption{Counts of large suspicious quasi-cliques, of sizes 7, 8, 9, 10, 11 and 12 found in select (k, d)-graphs\textsuperscript{\ref{tablenote1}}. Suspicious quasi-cliques highlighted in red.}
\label{table:quasicliqueresults}
\end{table}
\footnotetext{\label{tablenote1}Abbreviated table---some $(k, d)$-graphs not displayed.}

\noindent\textbf{(Quasi-)clique extraction.}
We extract cliques and quasi-cliques from the set of $(k,d)$-graphs. Cliques are complete sub-graphs of undirected graphs. Quasi-cliques are sub-graphs with edge densities\footnote{Number of edges that exist in sub-graph over number of edges in a complete graph with same number of vertices.} no less than a fixed threshold\footnote{$\theta = 0.90$}\cite{Uno}.

Identifying cliques is NP-hard. The Bron-Kerbosch algorithm finds maximal cliques and is based on the Branch-and-Bound technique. Most real-world datasets produce sparse graph, allowing Bron-Kerbosch to find maximal cliques faster than the theoretical worst case bound \cite{bron1973algorithm}. 
Suspiciously, large cliques of up to size 11 were found in the Yelp dataset (see Table \ref{table:cliqueresults}). Larger cliques with higher $k$ (more venues) and lower $d$ values (tighter time bound) are more suspicious.

To extract quasi-cliques, we utilize the method proposed in \cite{Uno}, which uses a greedy way to grow a quasi-clique, along with a clever pruning technique.
Quasi-cliques of size 11 and 12 were found (see Table \ref{table:quasicliqueresults}).

We manually inspected some flagged users, and were surprised that they are \textit{Yelp Scouts}\footnote{A user's Yelp Scout status is determined from a badge on their profile} who are paid by Yelp to review venues in new areas. In the $(6,5)$-graph, 31\% of users were/are Yelp Scouts.

This is a significant discovery; no prior study has revealed how these Scouts operate, how they choose which venues to visit, and what kinds of reviews they write (positive or negative)?
Our work sheds light on these little-known activities, which are highly organized both in timing and in venue selection.
While they may not be suspicious, they are certainly unnatural, and possibly controversial!

\begin{figure}[t]
\centering
\includegraphics[width=0.8\linewidth]{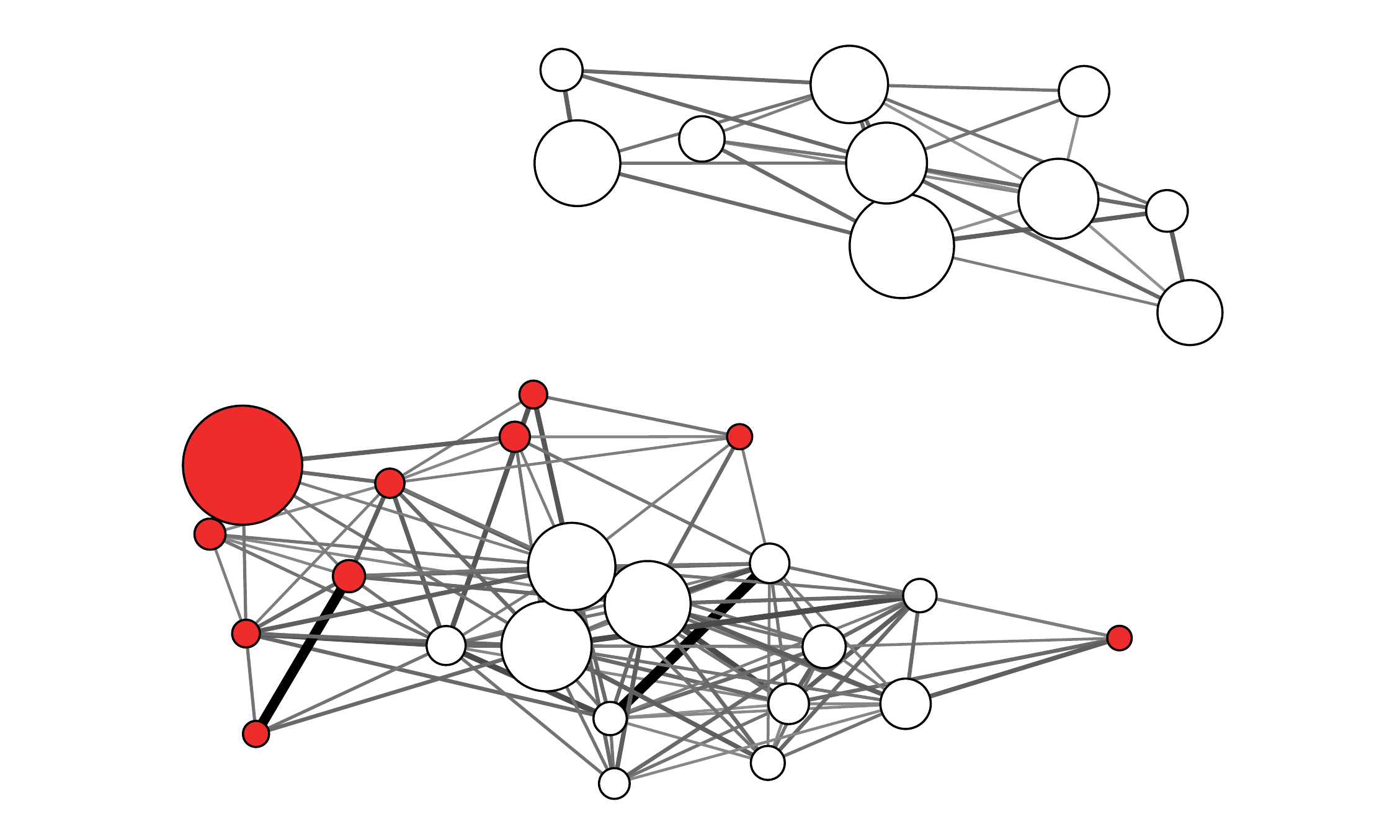}
\caption{Graph of combined cliques in a weighted $(6,5)$-graph (weighted by size of the intersection of friends). Larger nodes represent more reviews. Red nodes are Yelp Scouts while white nodes are regular users. Scouts are tightly clustered and appear to associate with other Scouts.}
\label{fig:cliquechart1}
\vskip-1em
\end{figure}

\section{Conclusions \& Next Steps}
It is alarming to find large cliques from the Yelp data, which are likely  suspicious.
But it is also critical to devise methods that reduce the false alarm rates to the minimum possible.

We plan to incorporate other rich signals from the Yelp data to help with this, such as by analyzing review text, and the spatial and temporal relationships among reviewed venues (e.g., it would be impossible for a user to visit a venue in the US and another in Asia on the same day).

\section*{Acknowledgment}
This work was supported in part by NSF grant CCF-1101283.

\bibliography{paper}
\bibliographystyle{IEEEtran}

\end{document}